\documentclass[11pt]{article}

\usepackage{amsmath}
\usepackage{amssymb}
\usepackage{bm}
\usepackage{xcolor}
\usepackage{graphicx}
\usepackage{float}
\usepackage{placeins}
\usepackage{hyperref}
\usepackage[margin=1.5in]{geometry}

\newcommand{\mcu}{\mathcal{U}}

\usepackage{draftwatermark}
\SetWatermarkText{\bf PREPRINT ~~~ PREPRINT ~~~ PREPRINT   ~~~ PREPRINT}
\SetWatermarkFontSize{1.5cm}
\SetWatermarkAngle{90}
\SetWatermarkHorCenter{20.5cm}
\SetWatermarkColor[gray]{0.75}

\title{Higher-order moments of the Mott-Smith shock approximation}

\author{Stefano Boccelli$^{1,2}$\\[1ex]$^1$NASA Goddard Space Flight Center\\$^2$NPP NASA Postdoctoral Fellow, Oak Ridge Associated Universities\\[1ex] email: \texttt{stefano.boccelli@nasa.gov}}
\date{2024}

\begin{document}

\maketitle

\begin{abstract}
This technical note reports the expression of selected higher-order moments associated with the Mott-Smith solution of the shock-wave profile.
The considered moments are the pressure tensor, the heat-flux vector and tensor, 
the fourth-order double-tensor, its full contraction, and the fifth-order moment vector.
The resulting shock profiles are shown for Mach 2 and Mach 10 conditions.
\end{abstract}



\section{Introduction}\label{sec:introduction}

The internal structure of a shock wave cannot be accurately described by the continuum theory.
Instead, one should employ the kinetic theory of gases \cite{ferziger1972mathematical} that describes the evolution of the velocity distribution function (VDF)
in phase space.
The Mott-Smith method allows for a simple closed-form solution of this problem \cite{mott1951solution}.
Despite its approximated nature and despite the present availability of accurate Monte-Carlo simulation tools,
this method remains of interest due to its simplicity and widespread adoption \cite{solovchuk2010prediction,timokhin2022mott,bret2024width,yilmaz2024nonlinear}.

The original work by Mott-Smith describes the shock-wave structure in terms of density, velocity and temperature profiles.
Later works \cite{nathenson1973constitutive,timokhin2022mott} also report the expression of the heat-flux vector and tensor.
In this technical note, the analysis is extended to higher-order moments of the Mott-Smith VDF.
These moments are meaningful non-equilibrium indicators, 
and are of interest to the development of moment methods \cite{muller1993extended,torrilhon2016modeling,timokhin2021local}.

\subsection{The Mott-Smith approximation}

This section briefly recaps the Mott-Smith approximation \cite{mott1951solution}. 
The velocity distribution function (VDF) is assumed to be composed of two Maxwellians,
\begin{equation}
  f = f_\alpha + f_\beta \, .
\end{equation}

\noindent Unless specified otherwise, in this work Greek subscripts identify the populations that compose the VDF
while Roman subscripts denote the spacial components of a vector or a tensor.

The bulk velocity and temperature of the first population, $u_\alpha$ and $T_\alpha$, are set to the free-stream values, 
while $u_\beta$ and $T_\beta$ refer to the post-shock gas state, and are found from the Rankine-Huguniot jump relations.
The additional constraint necessary to close the system is obtained by considering either second or third-order moments of the collision operator.
Ultimately, the densities of the two populations are found to follow the relation
\begin{equation}\label{eq:rho-MS-sol}
  \begin{cases}
    \rho_\alpha/\rho_{\mathrm{FS}} = 1/(1 + e^{B \, x/\lambda_\mathrm{FS}}) \, , \\ 
    \rho_\beta/\rho_{\mathrm{FS}} =  (u_\alpha/u_\beta)/(1 + e^{-B \, x/\lambda_\mathrm{FS}}) \, ,
  \end{cases}
\end{equation}

\noindent where the physical position, $x$, is scaled with the free-stream mean-free-path, $\lambda_\mathrm{FS}$.
At $x = 0$, one has $\rho_\alpha = \rho_{\mathrm{FS}}/2$.
The parameter $B$, appearing in Eq.~\eqref{eq:rho-MS-sol}, is a function of the Mach number and its values are tabulated in the original work 
for selected molecular interaction potentials.
The choice of the potential and of the moment employed to close the system are known to have an impact on the solution \cite{muckenfuss1962some,ziering1961mean}.
However, this does not affect the derivations shown in this work.
Notice that the results of this work also do not depend on the specific expression of the mean free path, but only on the dimensionless quantity $x/\lambda_\mathrm{FS}$, 
that is here taken to be a single variable.

Considering an ideal gas, the pressure of the two populations is $P_\alpha = \rho_\alpha (k_\mathrm{B}/m) T_\alpha$, where $k_\mathrm{B}$ and $m$ are the Boltzmann constant and the molecular mass of the considered gas, and the same applies to $\beta$.
The gas density, bulk velocity and pressure along the shock profile are%
\footnote{The original publication includes a typo in the temperature equation: $\rho_\alpha T_\alpha$ and $\rho_\beta T_\beta$ are subtracted, but should be added instead.}
\begin{equation}\label{eq:sol-MS-rho-u-T}
  \begin{cases}
    \rho = \rho_\alpha + \rho_\beta \, , \\
    u    = (\rho_\alpha u_\alpha + \rho_\beta u_\beta)/\rho \, , \\
    T = (\rho_\alpha T_\alpha + \rho_\beta T_\beta)/\rho + \tfrac{m}{3 k_\mathrm{B}}(\rho_\alpha \rho_\beta/\rho^2)(u_\alpha - u_\beta)^2 \, ,
  \end{cases}
\end{equation}

\noindent and the pressure is $P = \rho (k_\mathrm{B}/m) T$.


\subsection{Moments of the distribution function}

The statistical moments of a VDF, $f$, are defined as \cite{ferziger1972mathematical}
\begin{equation}\label{eq:definition-moments-U}
  U_\psi = \iiint_{-\infty}^{+\infty} \psi \, f \, \mathrm{d}^3 v \equiv \left< \psi \right> \, ,
\end{equation}

\noindent where $\psi$ is a particle quantity.
Choosing the particle mass, momentum or energy, $\psi = \{m, mv_i, mv^2/2\}$, results in the 
mass, momentum and energy densities of the whole gas, 
\begin{equation}\label{eq:rho-ui-E}
  \rho    = \left< m \right> \ , \ \ 
  \rho u_i = \left< m v_i \right> \ , \ \ 
  \rho \varepsilon = \frac{1}{2} \left< m v^2\right> \, ,
\end{equation}

\noindent whose expressions, in the Mott-Smith approximation, are obtained from Eq.~\eqref{eq:sol-MS-rho-u-T}.
In this work, we are also interested in the pressure tensor, $P_{ij}$, 
and in the heat-flux vector, $q_i$,
\begin{equation}\label{eq:def-Pij-qi}
  P_{ij} = \left< m c_i c_j \right> \, , \ \ 
  q_{i}  = \frac{1}{2}\left< m c_i c_j c_j \right> \, ,
\end{equation}

\noindent where $c_i = v_i - u_i$ is the peculiar velocity and where repeated indices imply summation.
The hydrostatic pressure, $P$, is obtained from the trace of the pressure tensor: $P = P_{ii}/3$.
Also, we are interested in the following moments:
the full heat-flux tensor, $Q_{ijk}$, the partially-contracted fourth-order moment tensor, $R_{ijkk}$,
\begin{equation}
  Q_{ijk}  = \left< m c_i c_j c_k \right> \, , \ \ 
  R_{ijkk} = \left< m c_i c_j c_k c_k \right> \, , \ \ 
\end{equation}

\noindent and the contracted fifth-order moment, $S_{ijjkk}$,
\begin{equation}
  S_{ijjkk} = \left< m c_i c_j c_j c_k c_k \right> \, .
\end{equation}

The fully-contracted fourth-order moment, $R_{iijj}$, is proportional to the kurthosis of the distribution function.
To fix the ideas, a super-Maxwellian kurthosis is often associated with heavier-than-Maxwellian tails.
At equilibrium, for a Maxwellian distribution, one has $Q_{ijk}^\mathrm{eq} = S^\mathrm{eq}_{ijjkk} = 0$, 
while the fourth-order moments are $R^\mathrm{eq}_{ijkk} = 5 P^2 \delta_{ij}/\rho$ (where $\delta_{ij}$ is the Kronecker delta), and $R^\mathrm{eq}_{iijj} = 15 P^2/\rho$.
In this technical note, the analysis is limited to the mentioned moments, as they represent the first extension to the Euler and Navier-Stokes theories,
and are commonly employed in third- and fourth-order moment methods \cite{mcdonald2013affordable,struchtrup2003regularization,alvarez2022regularized,boccelli2023modeling}.

\section{Moments of the Mott-Smith approximation}

The first three moments of the Mott-Smith approximation, $\rho$, $u$ and $P$, are obtained from Eq.~\eqref{eq:rho-MS-sol}.
The computation of other moments is particularly simple, as the integrals of Eq.~\eqref{eq:definition-moments-U} 
apply separately to the two Maxwellians, $f_\alpha$ and $f_\beta$.
The full pressure tensor is obtained as 
\begin{equation}
  P_{ij} = \left< m c_i c_j \right>_\alpha + \left< m c_i c_j \right>_\beta  \, .
\end{equation}

Considering the first integral, $\left< \cdots \right>_\alpha$, the calculation is simplified by rewriting the 
peculiar velocity as 
\begin{equation}\label{eq:shift-alpha-peculiar-vel}
  c_i \equiv v_i - u_i = c_{i, \alpha} + \mcu_{i, \alpha} \, ,
\end{equation}

\noindent where $\mcu_{i, \alpha} = u_{i,\alpha} - u_i$ expresses the deviation of the average velocity of $f_\alpha$ from
the bulk velocity of the whole distribution, $f$.
Notice that $\mcu_{i,\alpha}$ has only one non-zero component, $\mcu_{i,\alpha} = \mcu_\alpha \delta_{ix} = (u_\alpha - u)\delta_{ix}$.
This quantity is known along the shock profile from the Mott-Smith solution.
Also, the quantity $c_{i,\alpha}$ is the peculiar velocity with respect to the bulk velocity of the $\alpha$ species, $c_{i,\alpha} = v - u_{i\alpha}$, and 
can be used to easily compute the central moments associated with the $\alpha$ population.
With these definitions,
\begin{multline}\label{eq:calculation-Pij}
  \left< m c_i c_j \right>_\alpha 
= m \left< (c_{i,\alpha} + \mcu_{i,\alpha}) (c_{j,\alpha} + \mcu_{j,\alpha})\right>_\alpha \\
= m \left< c_{i,\alpha} c_{j,\alpha} + \mcu_{i,\alpha} c_{j,\alpha} + c_{i,\alpha} \mcu_{j,\alpha} + \mcu_{i,\alpha}\mcu_{j,\alpha}
    \right>_\alpha \\
  = P_{ij,\alpha} + \rho_\alpha \mcu_{i,\alpha} \mcu_{j,\alpha} \, ,
\end{multline}

\noindent where the terms involving odd-order powers of $c_{i,\alpha}$ are zero, since $f_\alpha$ is symmetric.
Furthermore, one considers that the pressure tensor, $P_{ij,\alpha}$, is isotropic. 
Repeating the computation for $f_\beta$ and adding the two contributions, one has
\begin{equation}
  P_{ij} =   (P_\alpha + P_\beta)\delta_{ij} + \left( \rho_\alpha \mcu_\alpha^2 + \rho_\beta \mcu_\beta^2 \right)\delta_{ix}\delta_{jx} \, ,
\end{equation}

\noindent or,
\begin{equation}\label{eq:result-Pxx-Pyy}
  \begin{cases}
    P_{xx} = P_\alpha + P_\beta + \rho_\alpha (u_\alpha - u)^2 + \rho_\beta (u_\beta - u)^2 \, , \\
    P_{yy} = P_{zz} = P_\alpha + P_\beta \, ,
  \end{cases}
\end{equation}

\noindent the remaining components being zero.
As expected, this result shows anisotropy, as $P_{xx} \ge P_{yy} = P_{zz}$.
In the free-stream and in the post-shock regions isotropy is recovered as either one of the two densities, or the respective velocity deviation, goes to zero.

The Mott-Smith heat-flux vector, $q_{i} = Q_{ijj}/2$, and the component $Q_{xxx}$ of the heat-flux tensor, are discussed in \cite{nathenson1973constitutive}.
Here, we report the expression of the full $Q_{ijk}$.
The derivation is analogous to that of the pressure tensor (Eq.~\eqref{eq:calculation-Pij}), and gives
\begin{equation}
  Q_{ijk} = \left( \mcu_\alpha P_\alpha + \mcu_\beta P_\beta \right) \left( \delta_{ij} \delta_{kx} + \delta_{ik} \delta_{jx} + \delta_{jk}\delta_{ix} \right)
          + \left( \rho_\alpha \mcu_\alpha^3 + \rho_\beta \mcu_\beta^3\right) \delta_{ix} \delta_{jx} \delta_{kx} \, .
\end{equation}

The heat-flux vector is found by contracting $Q_{ijk}$. 
As expected from symmetry considerations, its only non-zero component is $q_x$, which reads
\begin{equation}\label{eq:result-qx}
  q_x = \frac{Q_{xjj}}{2} = \frac{1}{2}\left[ \rho_\alpha \mcu_\alpha^3 + \rho_\beta \mcu_\beta^3 + 5 \, \mcu_\alpha P_\alpha + 5 \, \mcu_\beta P_\beta  \right] \, .
\end{equation}

With analogous computations, the fourth-order moment, $R_{ijkk}$, is found to be 
\begin{multline}
  R_{ijkk} = \left[ 5 P_\alpha^2/\rho_\alpha + \mcu_\alpha^2 P_\alpha + 5 P_\beta^2/\rho_\beta + \mcu_\beta^2 P_\beta \right] \delta_{ij} \\
           + \left[ \rho_\alpha \mcu_\alpha^4 + 7 \, \mcu_\alpha^2 P_\alpha + \rho_\beta \mcu_\beta^4 + 7 \, \mcu_\beta^2 P_\beta \right] \delta_{ix}\delta_{jx}    \, ,
\end{multline}

\noindent that is composed of an isotropic part, augmented by an additional contribution in the $(x,x)$ direction.
Its contraction, the scalar quantity $R_{iijj}$, is
\begin{equation}\label{eq:result-Riijj}
  R_{iijj} = \rho_\alpha \mcu_\alpha^4 + 10 \, \mcu_\alpha^2 P_\alpha + 15 P_\alpha^2/\rho_\alpha 
             + \rho_\beta \mcu_\beta^4 + 10 \, \mcu_\beta^2 P_\beta + 15 P_\beta^2/\rho_\beta \, .
\end{equation}

Similarly, one can find the contracted fifth-order moment vector, whose only non-zero component is 
\begin{equation}\label{eq:result-Sxjjkk}
  S_{xjjkk} = \rho_\alpha \mcu_\alpha^5 + 14 \, \mcu_\alpha^3 P_\alpha + 35 \, \mcu_\alpha P_\alpha^2/\rho_\alpha 
            + \rho_\beta \mcu_\beta^5 + 14 \, \mcu_\beta^3 P_\beta + 35 \, \mcu_\beta P_\beta^2 / \rho_\beta \, .
\end{equation}

\subsubsection*{Extension to N populations}

Extensions of the Mott-Smith theory to a higher number of Maxwellian populations have been proposed in the past.
For instance, Salwen et al \cite{salwen1964extension} considered a distribution in the form $f = \sum_{\gamma=1}^N f_\gamma$, 
with $N=3$ distinct populations.
The expressions for the moments, discussed in this work, are easily extended to these cases by simply adding the contributions of the individual populations.
For instance, the pressure tensor would read
\begin{equation}
  P_{ij} = \sum_{\gamma = 1}^{N} \left[ P_\gamma \delta_{ij} + \rho_\gamma (u_\gamma - u)^2\delta_{ix}\delta_{jx} \right] \, ,
\end{equation}

\noindent and the other moments are obtained similarly.


\section{Numerical computations}

This section shows the shock structure for two Mach numbers, $M=2$ and $M=10$.
These conditions are chosen arbitrarily, and are intended to span a large range of velocities.
Above $M=10$, the density profile becomes substantially insensitive to the free-stream velocity, as the parameter $B$ changes only marginally \cite{mott1951solution}.

\begin{figure*}
  \centering
  \includegraphics[width=\textwidth]{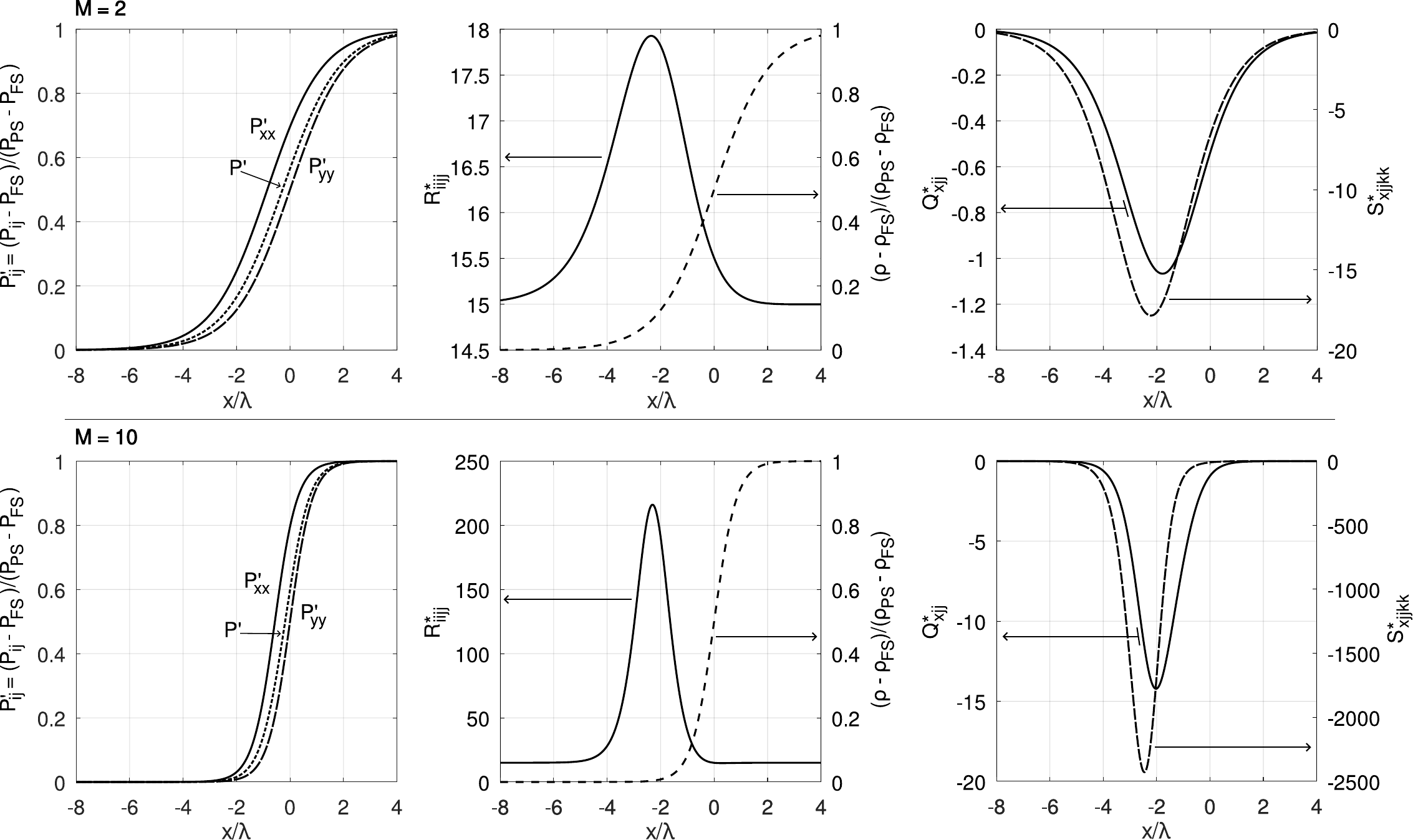}
  \caption{Shock wave profiles from the Mott-Smith model (with $B=B_1$, see \cite{mott1951solution}) for $M = 2$ (Top) and $M = 10$ (Bottom).} 
  \label{fig:MS-profile-moments}
\end{figure*}

First, the Mott-Smith solution is computed in terms of density, velocity and pressure, as detailed above.
Figure~\ref{fig:MS-profile-moments}-Left and -Center show the pressure and density profiles, rescaled with the free-stream and post-shock values.
In these calculations, the parameter $B$ was set to the value $B_1$ tabulated in the original work.
The expressions of the higher-order moments remain valid for other choices of $B$.

Figure~\ref{fig:MS-profile-moments}-Left also shows the pressure tensor components, $P_{xx}$ and $P_{yy}$, obtained from Eq.~\eqref{eq:result-Pxx-Pyy},
while the Center and Right boxes show the dimensionless moments, $R_{iijj}^\star$, $Q_{xjj}^\star$ and $S_{xjjkk}^\star$, 
from Eqs.~\eqref{eq:result-Riijj}, \eqref{eq:result-qx} and \eqref{eq:result-Sxjjkk}, respectively.
The non-dimensionalization (superscript ``$^\star$'') is performed dividing the moments by the local density, $\rho$, and by suitable powers of the 
thermal velocity, $\sqrt{P/\rho}$,
\begin{equation}
  \begin{cases}
    Q_{xjj}^\star = Q_{xjj}/\left[\rho \, (P/\rho)^{3/2}\right] \, , \\
    R_{iijj}^\star = R_{iijj}/\left[\rho \, (P/\rho)^{4/2}\right] \, , \\
    S_{xjjkk}^\star = S_{xjjkk}/\left[\rho \, (P/\rho)^{5/2}\right] \, .
  \end{cases}
\end{equation}

At equilibrium, $R_{iijj}^{\star \, \mathrm{eq}} = 15$ and $Q_{xjj}^{\star \, \mathrm{eq}} =S_{xjjkk}^{\star \, \mathrm{eq}} = 0$.
As one might expect, these moments depart significantly from equilibrium in the $M=2$ shock, and reach extreme values in the $M = 10$ case.

The maximum of these moments is located upstream of the shock.
This behavior of the higher-order moments is compatible with the previous observation by Muckenfuss \cite{muckenfuss1962some} that the center of the temperature profile is located ahead
of the center of the velocity profile, which in turn is ahead of the center of the density profile (located at $x=0$).
Although not shown directly in the figure, the heat-flux is located upstream of the temperature profile, and moments of higher order are generally further upstream.
Notice that this phenomenon is well-known \cite{mcdonald2013affordable,laplante2016comparison}, and is due to the increased sensitivity of higher-order moments to supra-thermal particles.

\section*{Data availability statement}
This manuscript has no associated data.

\section*{Acknowledgements}
This research was partially supported by an NPP (NASA Postdoctroal Program) appointment, administered by Oak Ridge Associated Universities (ORAU).

%

\end{document}